# Detect adverse drug reactions for drug Simvastatin

Yihui Liu[1,1.1.1 1 2]

[1]Institute of Intelligent Information Processing
Shandong Polytechnic University, China
UK Yihui_liu_2005@yahoo.co.uk

Uwe Aickelin[2]

[2]Department of Computer Science,
University of Nottingham,

*Abstract*—Adverse drug reaction (ADR) is widely concerned for public health issue. In this study we propose an original approach to detect the ADRs using feature matrix and feature selection. The experiments are carried out on the drug Simvastatin. Major side effects for the drug are detected and better performance is achieved compared to other computerized methods. The detected ADRs are based on the computerized method, further investigation is needed.

## I. INTRODUCTION

Adverse drug reaction (ADR) is widely concerned for public health issue. ADRs are one of most common causes to withdraw some drugs from market [1]. Now two major methods for detecting ADRs are spontaneous reporting system (SRS) [2, 3], and prescription event monitoring (PEM) [4, 5]. The World Health Organization (WHO) defines a signal in pharmacovigilance as "any reported information on a possible causal relationship between an adverse event and a drug, the relationship being unknown or incompletely documented previously"[6]. For spontaneous reporting system, many machine learning methods are used to detect ADRs, such as Bayesian confidence propagation neural network (BCPNN) [7], decision support method [8], genetic algorithm [9], knowledge based approach [10], etc. One limitation is the reporting mechanism to submit ADR reports [8], which has serious underreporting and is not able to accurately quantify the corresponding risk. Another limitation is hard to detect ADRs with small number of occurrences of each drug-event association in the database.

In this paper we propose feature selection approach to detect ADRs from The Health Improvement Network (THIN) database. First feature matrix, which represents the medical events for the patients before and after taking drugs, is created by linking patients' prescriptions and corresponding medical events together. Then significant features are selected based on feature selection methods, comparing the feature matrix before patients take drugs with one after patients take drugs. Finally the significant ADRs can be detected from thousands of medical events based on corresponding features. Experiments are carried out on the drug Simvastatin. Good performance is achieved.

## II. FEATURE MATRIX AND FEATURE SELECTION

### A. The Extraction of Feature Matrix

To detect the ADRs of drugs, first feature matrix is extracted from THIN database, which describes the medical events that patients occur before or after taking drugs. Then feature selection method of Student's t-test is performed to select the significant features from feature matrix containing thousands of medical events. Figure 1 shows the process to detect the ADRs using feature matrix. Feature matrix $A$ describes the medical events for each patient during 60 days before they take drugs. Feature matrix $B$ reflects the medical events during 60 days after patients take drugs. In order to reduce the effect of the small events, and save the computation time and space, we set 100 patients as a group. Matrix $X$ and $Y$ are feature matrix after patients are divided into groups.

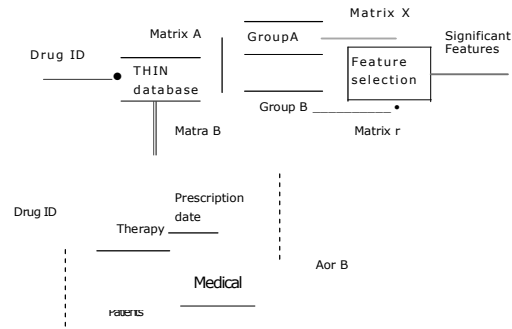

Figure 1. The process to detect ADRs. Matrix $A$ and $B$ are feature matrix before patients take drugs or after patients take drugs. The time period of observation is set to 60 days. Matrix $X$ and $Y$ are feature matrix after patients are divided into groups. We set 100 patients as one group.

### B. Medical Events and Readcodes

Medical events or symptoms are represented by medical codes or Readcodes. There are 103387 types of medical events in "Readcodes" database. The Read Codes used in general practice (GP), were invented and developed by Dr James Read in 1982. The NHS (National Health Service) has expanded the codes to cover all areas of clinical practice. The code is hierarchical from left to right or from level 1 to level 5. It means that it gives more detailed information from level 1 to level 5. Table 1 shows the medical symptoms based on Readcodes at level 3 and at level 5. 'Other soft tissue disorders' is general term using Readcodes at level 3. `Foot pain', 'Heel pain', etc., give more details using Readcodes at level 5.

## C. Feature Selection Based on Student's t-test

Feature extraction and feature selection are widely used in biomedical data processing [11-18]. In our research we use Student's t-test [19] feature selection method to detect the significant ADRs from thousands of medical events. Student's t-test is a kind of statistical hypothesis test based on a normal distribution, and is used to measure the difference between two kinds of samples.

TABLE I.
MEDICAL EVENTS BASED ON READCODES AT LEVEL 3 AND LEVEL 5.

| | Level | Readcodes | Medical events |
|---|---|---|---|
| Muscle pain | Level 3 | N24..00 | Other soft tissue disorders |
| | Level 5 | N245.16 | Leg pain |
| | | N245111 | Toe pain |
| | | N245.13 | Foot pain |
| | | N245700 | Shoulder pain |
| | | N245.15 | Heel pain |

## D. Other Parameters

The variable of ratio $R$, is defined to evaluate significant changes of the medical events, using ratio of the patient number after taking the drug to one before taking the drug. The variable $R_2$ represents the ratio of patient number after taking the drug to the number of whole population having one particular medical symptom.

The ratio variables $R_1$ and $R_2$ are defined as follows:

$$R_1 = \begin{cases} N_A / N_B & \text{if } N_B \neq 0; \\ N_A & \text{if } N_B = 0; \end{cases}$$

$$R_2 = N_A / N$$

where $N_B$ and $N_A$ represent the numbers of patients before or after they take drugs for having one particular medical event respectively. The variable $N$ represents the number of whole population who take drugs.

## III. EXPERIMENTS AND RESULTS

Simvastatin [20], under the trade name Zocor, is a hypolipidemic drug used to control elevated cholesterol, or hypercholesterolemia. It is a member of the statin class of pharmaceuticals. Simvastatin has side effects [20,21,22]: severe allergic reactions (rash; hives; itching; difficulty breathing; tightness in the chest; swelling of the mouth, face, lips, or tongue; unusual hoarseness); burning, numbness, or tingling; change in the amount of urine produced; confusion; dark or red-colored urine; decreased sexual ability; depression; dizziness; fast or irregular heartbeat; fever, chills, or persistent sore throat; joint pain; loss of appetite; memory problems; muscle pain, tenderness, or weakness (with or without fever and fatigue); pale stools; red, swollen, blistered, or peeling skin; severe or persistent nausea or stomach or back pain; shortness of breath; trouble sleeping; unusual bruising or bleeding; unusual tiredness or weakness; vomiting; yellowing of the skin or eyes.

14905 patients from 20GP data in THIN database are taking Simvastatin, and 13060 medical events are obtained based on Readcodes at level 1-5. After grouping them, 149x13060 feature matrix is obtained. For Readcodes at level 1-3, 149x2693 feature matrix is obtained.

Table 2 shows the top 30 detected results in ascending order of p value of Student's t-test, using Readcodes at level 1-5 and at level 1-3. The detected results are using p value less than 0.05, which represent the significant change after patients take the drug. Table 3 shows the results in descending order of the ratio of the number of patients after taking the drug to one before taking the drug. Table 4 shows potential ADRs related cancer for Simvastatin. The detected ADRs are based on our computerized method, further investigation is needed.

It is clear that our detected results are consistent with published side effects for statin drugs [21, 22]. Major ADRs of 'muscle and musculoskeletal' events for statin drugs are detected not only based on Readcodes at level 1-5, but also based on Readcodes at level 1-3.

## IV. CONCLUSIONS

In this study we propose a novel method to successfully detect the ADRs using feature matrix and feature selection. A feature matrix, which characterizes the medical events before patients take drugs or after patients take drugs, is created from THIN database. The feature selection method of Student's t-test is used to detect the significant features from thousands of medical events. The significant ADRs, which are corresponding to significant features, are detected. Experiments are performed on the drug Simvastatin. Compared to other computerized method, our proposed method achieves good performance.

TABLE II. THE TOP 20 ADRS FOR SIMVASTATIN BASED ON P VALUE OF STUDENT'S T-TEST.

| | Rank | Readcodes | Medical events | NB | NA | R1 | R2 |
|---|---|---|---|---|---|---|---|
| Level 1-5 | 1 | 1Z12.00 | Chronic kidney disease stage 3 | 185 | 1095 | 5.92 | 7.35 |
| | 2 | M03z000 | C,ellulitis NOS | 98 | 503 | 5.13 | 3.37 |
| | 3 | F4C0.00 | Acute conjunctivitis | 113 | 525 | 4.65 | 3.52 |
| | 4 | N131.00 | Cervicalgia - pain in neck | 140 | 609 | 4.35 | 4.09 |
| | 5 | 1106z000 | Chest infection NOS | 284 | 1201 | 4.23 | 8.06 |
| | 6 | N143.00 | Sciatica | 83 | 366 | 4.41 | 2.46 |
| | 7 | F46..00 | Cataract | 40 | 312 | 7.80 | 2.09 |
| | 8 | 1M10.00 | Knee pain | 198 | 762 | 3.85 | 5.11 |
| | 9 | A53..11 | Shingles | 41 | 262 | 6.39 | 1.76 |
| | 10 | C34..00 | Gout | 107 | 381 | 3.56 | 2.56 |
| | 11 | 1A55.00 | Dysuria | 70 | 308 | 4.40 | 2.07 |
| | 12 | N245.17 | Shoulder pain | 185 | 717 | 3.88 | 4.81 |
| | 13 | F45..00 | Glaucoma | 20 | 148 | 7.40 | 0.99 |
| | 14 | K190.00 | Urinary tract infection, site not specified | 128 | 607 | 4.74 | 4.07 |
| | 15 | F501.00 | Infective otitis extema | 89 | 372 | 4.18 | 2.50 |
| | 16 | 1D14.00 | C/O: a rash | 152 | 689 | 4.53 | 4.62 |
| | 17 | N094K12 | Hip pain | 96 | 461 | 4.80 | 3.09 |
| | 18 | 1832.11 | Ankle swelling symptom | 34 | 190 | 5.59 | 1.27 |
| | 19 | 1C9..00 | Sore throat symptom | 97 | 410 | 4.23 | 2.75 |
| | 20 | B33..11 | Basal cell carcinoma | 42 | 212 | 5.05 | 1.42 |
| Level 1-3 | 1 | 1106..00 | Acute bronchitis and bronchiolitis | 598 | 2221 | 3.71 | 14.90 |
| | 2 | 1Z1..00 | Chronic renal impairment | 213 | 1286 | 6.04 | 8.63 |
| | 3 | 171..00 | Cough | 571 | 2192 | 3.84 | 14.71 |
| | 4 | N24..00 | Other soft tissue disorders | 807 | 2643 | 3.28 | 17.73 |
| | 5 | N21..00 | Peripheral enthesopathies and allied syndromes | 265 | 1054 | 3.98 | 7.07 |
| | 6 | 1105..00 | Other acute upper respiratory infections | 213 | 1074 | 5.04 | 7.21 |
| | 7 | M03..00 | Other cellulitis and abscess | 140 | 659 | 4.71 | 4.42 |
| | 8 | F4C..00 | Disorders of conjunctiva | 147 | 731 | 4.97 | 4.90 |
| | 9 | 173..00 | Breathlessness | 461 | 1403 | 3.04 | 9.41 |
| | 10 | 19F..00 | Diarrhoea symptoms | 189 | 861 | 4.56 | 5.78 |
| | 11 | K19..00 | Other urethral and urinary tract disorders | 221 | 1010 | 4.57 | 6.78 |
| | 12 | 183..00 | Oedema | 177 | 795 | 4.49 | 5.33 |
| | 13 | N09..00 | Other and unspecified joint disorders | 355 | 1413 | 3.98 | 9.48 |
| | 14 | N13..00 | Other cervical disorders | 146 | 648 | 4.44 | 4.35 |
| | 15 | F46..00 | Cataract | 67 | 435 | 6.49 | 2.92 |
| | 16 | 1B1..00 | General nervous symptoms | 437 | 1413 | 3.23 | 9.48 |
| | 17 | J57..00 | Other disorders of intestine | 75 | 361 | 4.81 | 2.42 |
| | 18 | N14..00 | Other and unspecified back disorders | 246 | 984 | 4.00 | 6.60 |
| | 19 | 1M1..00 | Pain in lower limb | 228 | 851 | 3.73 | 5.71 |
| | 20 | 1D1..00 | GO: a general symptom | 317 | 1278 | 4.03 | 8.57 |

Variable NB and NA represent the numbers of patients before or after they take drugs for having one particular medical event Variable R1 represents the ratio of the numbers of patients after taking drugs to the numbers of patients before taking drugs. Variable R2 represents the ratio of the numbers of patients after taking drugs to the number of the whole population.

TABLE III. THE TOP 20 ADRS FOR SIMVASTATIN BASED ON DESCENDING ORDER OF R1 VALUE.

| | Rank | Readcodes | Medical events | NB | NA | R1 | R2 |
|---|---|---|---|---|---|---|---|
| Level 1-5 | 1 | 1Z1E.00 | Chronic kidney disease stage 3A without proteinuria | 0 | 40 | 40.00 | 0.27 |
| | 2 | Eu32000 | [X]Mild depressive episode | 1 | 39 | 39.00 | 0.26 |
| | 3 | C106.00 | Diabetes mellitus with neurological manifestation | 1 | 39 | 39.00 | 0.26 |
| | 4 | Eu32100 | [X]Moderate depressive episode | 0 | 27 | 27.00 | 0.18 |
| | 5 | SK17100 | Other leg injury | 1 | 26 | 26.00 | 0.17 |
| | 6 | 11120.11 | Catarrh unspecified | 0 | 26 | 26.00 | 0.17 |
| | 7 | S646000 | Minor head injury | 2 | 51 | 25.50 | 0.34 |
| | 8 | 1125..00 | Bronchopneumonia due to unspecified organism | 1 | 24 | 24.00 | 0.16 |
| | 9 | 173L.00 | MRC Breathlessness Scale: grade 5 | 0 | 22 | 22.00 | 0.15 |

|   | 10 | 16D1.00 | Recurrent falls | 0 | 21 | 21.00 | 0.14 |
|---|----|---------|-----------------|---|----|-------|------|
|   | 11 | B32..00 | Malignant melanoma of skin | 1 | 20 | 20.00 | 0.13 |
|   | 12 | E290000 | Grief reaction | 1 | 20 | 20.00 | 0.13 |
|   | 13 | J4z0.00 | Non-infective gastritis NOS | 1 | 20 | 20.00 | 0.13 |
|   | 14 | SN15.00 | Chilblains | 1 | 20 | 20.00 | 0.13 |
|   | 15 | M220.00 | Cutaneous horn | 1 | 20 | 20.00 | 0.13 |
|   | 16 | 1B1E.00 | Hallucinations | 1 | 19 | 19.00 | 0.13 |
|   | 17 | 1151..00 | Pleurisy | 1 | 19 | 19.00 | 0.13 |
|   | 18 | SE46.00 | Traumatic haematoma | 0 | 19 | 19.00 | 0.13 |
|   | 19 | 1D12.00 | CIO: stiffness | 1 | 19 | 19.00 | 0.13 |
|   | 20 | B834.00 | Carcinoma in situ of prostate | 1 | 18 | 18.00 | 0.12 |
| Level 1-3 | 1 | 1125..00 | Bronchopneumonia due to unspecified organism | 1 | 24 | 24.00 | 0.16 |
|   | 2 | 161..00 | Appetite symptom | 1 | 23 | 23.00 | 0.15 |
|   | 3 | B32..00 | Malignant melanoma of skin | 1 | 22 | 22.00 | 0.15 |
|   | 4 | U60..00 | [X]Drugs/meds/biol subs caus adverse effects in therap use | 2 | 37 | 18.50 | 0.25 |
|   | 5 | J12..00 | Duodenal ulcer - (DU) | 2 | 37 | 18.50 | 0.25 |
|   | 6 | N....00 | Musculoskeletal and connective tissue diseases | 0 | 16 | 16.00 | 0.11 |
|   | 7 | D0...00 | Deficiency anemias | 2 | 31 | 15.50 | 0.21 |
|   | 8 | B57..00 | Secondary malig neop of respiratory and digestive systems | 0 | 15 | 15.00 | 0.10 |
|   | 9 | K23..00 | Hydrocele | 2 | 29 | 14.50 | 0.19 |
|   | 10 | F4...00 | Disorders of eye and adnexa | 5 | 65 | 13.00 | 0.44 |
|   | 11 | B17..00 | Malignant neoplasm of pancreas | 0 | 13 | 13.00 | 0.09 |
|   | 12 | 17...00 | Respiratory symptoms | 0 | 13 | 13.00 | 0.09 |
|   | 13 | G8y..00 | Other specified vein, lymphatic or other circulatory | 1 | 13 | 13.00 | 0.09 |
|   | 14 | C26..00 | Vitamin B-complex deficiency | 3 | 37 | 12.33 | 0.25 |
|   | 15 | S5z..00 | Sprains and strains NOS | 2 | 24 | 12.00 | 0.16 |
|   | 16 | SF3..00 | Crush injury, lower limb | 0 | 12 | 12.00 | 0.08 |
|   | 17 | B11..00 | Malignant neoplasm of stomach | 1 | 12 | 12.00 | 0.08 |
|   | 18 | H41..00 | Asbestosis | 1 | 12 | 12.00 | 0.08 |
|   | 19 | F43..00 | Chorioretinal inflammations scars and other disorders | 1 | 12 | 12.00 | 0.08 |
|   | 20 | F5...00 | Diseases of the ear and mastoid process | 1 | 12 | 12.00 | 0.08 |

TABLE W. THE POTENTIAL ADRS RELATED TO CANCER FOR SIMVASTATIN BASED ON P VALUE OF STUDENT'S T-TEST.

| Rank | Readcodes | Medical events | NB | NA | R1 | R2 |
|------|-----------|----------------|----|----|----|----|
| 1 | B33..00 | Other malignant neoplasm of skin | 46 | 241 | 5.24 | 1.62 |
| 2 | B34..00 | Malignant neoplasm of female breast | 8 | 72 | 9.00 | 0.48 |
| 3 | B76..00 | Benign neoplasm of skin | 75 | 240 | 3.20 | 1.61 |
| 4 | BB5..00 | [M]Adenomas and adenocarcinomas | 7 | 69 | 9.86 | 0.46 |
| 5 | BB2..00 | [M]Papillary and squamous cell neoplasms | 7 | 66 | 9.43 | 0.44 |
| 6 | B46..00 | Malignant neoplasm of prostate | 23 | 105 | 4.57 | 0.70 |
| 7 | B22..00 | Malignant neoplasm of trachea, bronchus and lung | 5 | 47 | 9.40 | 0.32 |
| 8 | B32..00 | Malignant melanoma of skin | 1 | 22 | 22.00 | 0.15 |
| 9 | 170..00 | Suspected malignancy | 4 | 33 | 8.25 | 0.22 |
| 10 | BB3..00 | [M]Basal cell neoplasms | 4 | 30 | 7.50 | 0.20 |
| 11 | B8...00 | Carcinoma in situ | 12 | 41 | 3.42 | 0.28 |
| 12 | B57..00 | Secondary malig neop of respiratory and digestive systems | 0 | 15 | 15.00 | 0.10 |
| 13 | B83..00 | Carcinoma in situ of breast and genitourinary system | 4 | 29 | 7.25 | 0.19 |
| 14 | B13..00 | Malignant neoplasm of colon | 7 | 30 | 4.29 | 0.20 |
| 15 | B17..00 | Malignant neoplasm of pancreas | 0 | 13 | 13.00 | 0.09 |
| 16 | B62..00 | Other malignant neoplasm of lymphoid and histiocytic tissue | 2 | 17 | 8.50 | 0.11 |
| 17 | B11..00 | Malignant neoplasm of stomach | 1 | 12 | 12.00 | 0.08 |
| 18 | B49..00 | Malignant neoplasm of urinary bladder | 3 | 19 | 6.33 | 0.13 |
| 19 | B14..00 | Malignant neoplasm of rectum, rectosigmoid junction and anus | 5 | 24 | 4.80 | 0.16 |
| 20 | B63..00 | Multiple myeloma and immunoproliferative neoplasms | 0 | 8 | 8.00 | 0.05 |
| 21 | B59..00 | Malignant neoplasm of unspecified site | 1 | 11 | 11.00 | 0.07 |
| 22 | B10..00 | Malignant neoplasm of oesophagus | 3 | 12 | 4.00 | 0.08 |
| 23 | BB4..00 | [M]Transitional cell papillomas and carcinomas | 6 | 17 | 2.83 | 0.11 |
| 24 | B1 z..00 | Malig neop oth/ill-defined sites digestive tract/peritoneum | 0 | 5 | 5.00 | 0.03 |
| 25 | B58..00 | Secondary malignant neoplasm of other specified sites | 2 | 9 | 4.50 | 0.06 |
| 26 | B44..00 | Malignant neoplasm of ovary and other uterine adnexa | 0 | 4 | 4.00 | 0.03 |
| 27 | B81..00 | Carcinoma in situ of respiratory system | 0 | 4 | 4.00 | 0.03 |
| 28 | B56..00 | Secondary and unspecified malignant neoplasm of lymph | 0 | 4 | 4.00 | 0.03 |
| 29 | B71..00 | Benign neoplasm of other parts of digestive system | 10 | 23 | 2.30 | 0.15 |
| 30 | BBQ..00 | [M]Germ cell neoplasms | 0 | 3 | 3.00 | 0.02 |